\documentclass[12pt]{article}
\usepackage{epsfig,amssymb,amsmath,psfrag,subfigure,rotate}

\allowdisplaybreaks
\newcommand{\insertfig}[2]{\mbox{\epsfxsize=#1cm \epsfbox{#2.eps}}}

\def \be  {\begin{equation}}
\def \ee  {\end{equation}}
\def \ba  {\begin{eqnarray}}
\def \ea  {\end{eqnarray}}
\def \baa {\begin{eqnarray*}}
\def \eaa {\end{eqnarray*}}
\def \lab #1 {\label{#1}}

\def \matrix #1 {\left(\begin{array}{cc} #1 \end{array}\right)}

\def \tr {\mathop{\rm tr}\nolimits}

\newcommand \vev [1] {\langle{#1}\rangle}
\newcommand \VEV [1] {\left\langle{#1}\right\rangle}

\newcommand \bra [1] {\langle {#1}|}

\newcommand{\ft}[2]{{\textstyle\frac{#1}{#2}}}
\begin{document}

\begin{titlepage}

\thispagestyle{empty}

\vspace*{3cm}

\centerline{\large \bf A note on two-loop superloop}
\vspace*{1cm}

\centerline{\sc A.V. Belitsky}

\vspace{10mm}

\centerline{\it Department of Physics, Arizona State University}
\centerline{\it Tempe, AZ 85287-1504, USA}

\vspace{1cm}

\centerline{\bf Abstract}

\vspace{5mm}

We explore the duality between supersymmetric Wilson loop on null polygonal contours in maximally supersymmetric Yang-Mills theory and 
next-to-maximal helicity violating (NMHV) scattering amplitudes. Earlier analyses demonstrated that the use of a dimensional regulator for ultraviolet
divergences, induced due to presence of the cusps on the loop, yields anomalies that break both conformal symmetry and supersymmetry. At one-loop 
order, these are present only in Grassmann components localized in the vicinity of a single cusp and result in a universal function for any number of sites 
of the polygon that can be subtracted away in a systematic manner leaving a well-defined supersymmetric remainder dual to corresponding components 
of the superamplitude. The question remains though whether components which were free from the aforementioned supersymmetric anomaly at leading 
order of perturbation theory remain so once computed at higher orders. Presently we verify this fact by calculating a particular component of the null polygonal 
super Wilson loop at two loops restricting the contour kinematics to a two-dimensional subspace. This allows one to perform all computations in a concise 
analytical form and trace the pattern of cancellations between individual Feynman graphs in a transparent fashion. As a consequence of our consideration 
we obtain a dual conformally invariant result for the remainder function in agreement with one-loop NMHV amplitudes.

\end{titlepage}

\setcounter{footnote} 0

\newpage

\pagestyle{plain}
\setcounter{page} 1

The duality between scattering amplitudes in $\mathcal{N}=4$ super-Yang-Mills theory and a supersymmetric extension of the Wilson loop
spanned on a polygonal closed contour with its sites defined by particles' momenta involved in scattering occupied an important niche in
devising new techniques for analysis of dynamics of gauge theories at weak and strong coupling regimes and interpolation between the two. A 
distinguished role played by the maximal supersymmetry in four dimensions is that all particles of the theory can be combined into a single 
CPT self-conjugated light-cone superfield $\Phi$ defined by a (finite) series in the Grassmann variable $\eta^A$ with coefficients determined 
by the fields of appropriate helicity to compensate for the deficit introduced by the $\eta$ itself and matching $SU(4)$ tensor structure 
\cite{BriMan82,Nai88}. Thus, the $n-$particle S-matrix of the theory is concisely represented by the amputated Green $\mathcal{A}_n$ functions 
of $n$ superfields $\Phi$. Extracting the (super)momentum conservation laws allows one to cast the superamplitude $\mathcal{A}_n$ into the 
following form \cite{Nai88}
\be
\label{GenericA}
\mathcal{A}_n 
= 
i (2 \pi)^4 
\frac{
\delta^{(4)} \big( \sum\nolimits_i \lambda_i \tilde\lambda_i \big)  
\delta^{(8)} \big( \sum\nolimits_i \lambda_i \eta_i \big) 
}{\vev{12} \vev{23} \vev{34} \dots \vev{n-1 n}}
\widehat{\mathcal{A}}_n (\lambda_i, \widetilde\lambda_i, \eta_i)
\, .
\ee
The use of the spinor helicity formalism, adopted here and below, simplifies the representation of the amplitude. Namely, the massless particles' 
momenta  $p_i^{\dot\alpha\alpha} = \widetilde\lambda^{\dot\alpha} \lambda^\alpha$ and their chiral charges $q^{A, \alpha} = \eta^A \lambda^\alpha$ 
are written by means of commuting Weyl spinors $\lambda^\alpha$ and $\widetilde\lambda^{\dot\alpha}$, with their inner product 
defined as $\vev{ij} = \lambda^\alpha_i \lambda_{j \alpha}$ and $[ij] = \widetilde\lambda_{i \dot\alpha} \widetilde\lambda^{\dot\alpha}_j$, as well as 
anticommuting Grassmann variables $\eta^A$ transforming in the fundamental of $SU(4)$. The reduced amplitude $\widehat{\mathcal{A}}_n$ admits an 
expansion in terms of $\eta$'s
\be
\widehat{\mathcal{A}}_n = \widehat{\mathcal{A}}_{n, 0} + \widehat{\mathcal{A}}_{n, 1} + \dots
\, ,
\ee
that terminates at order\footnote{Nilpotence of the Grassmann variables alone is not sufficient to produce this constraint, the reduction by maximal degree
by four is a consequence of superconformal symmetry.} $k=n-4$, with each term being a homogeneous polynomial of degree $\eta^{4 k}$. Each term
in this expansion describes scattering of particle with total helicity $- n + 4 + 2k$, with the leading term being the maximal helicity-violating amplitudes (MHV),
then the next-to-maximal helicity-violating (NMHV) amplitude etc.  At tree level, $\widehat{\mathcal{A}}_{n, 0}^{(0)} = 1$, while the latter can be written as a sum 
\cite{DruHenKorSok10}
\be
\label{treeNMHV}
\widehat{\mathcal{A}}_{n, 1}^{(0)} =  \sum_{1< q < r < n} R_{n;qr}
\, ,
\ee 
of superconformal invariants
\be
R_{n;qr} 
= 
\frac{\delta^4 (\vev{n, q-1, q, r-1} \chi_r + \ {\rm cyclic})}{\vev{q-1, q, r-1, r} \vev{q, r-1, r, n} \vev{r-1, r, n, q-1} \vev{r, n, q-1, q} \vev{n, q-1, q, r-1}}
\, ,
\ee
written in terms of momentum twistors $Z^a_j = (\lambda^\alpha_j, x_j^{\dot\alpha\alpha} \lambda_{j \alpha})$, with angle-brackets being
$\vev{ijkl} = \varepsilon_{abcd} Z^a_i Z^b_j Z^c_k Z^d_l$. In what follows, the focus of our analysis  will be $\widehat{\mathcal{A}}_{n, 1}$ at higher
orders of perturbation theory.

A profound realization of the past few years was that the superamplitude (\ref{GenericA}) is expected to admit a dual representation in terms of a Wilson 
superloop \cite{ManSki10,Car10} spanned on a closed polygonal supercontour with its vertices localized at $(x_i, \theta^A_i)$ such that its path segments are 
proportional to the particles' (super)momenta, $p_i^{\dot\alpha \alpha} = (x_1 - x_{i+1})^{\dot\alpha \alpha} \equiv x_{i,i+1}^{\dot\alpha \alpha}$ and 
$q^{A \alpha} = (\theta_i - \theta_{i+1})^{A, \alpha} \equiv \theta_{i i+1}^{A, \alpha}$,
\be
\vev{\mathcal{W}_n (x_i, \theta_i)}
= 
\frac{1}{N_c} \VEV{\tr \left( \mathcal{W}_{[1n]} \dots \mathcal{W}_{[32]}  \mathcal{W}_{[21]} \right) }
\, ,
\ee
where
\be
\mathcal{W}_{[i+1, i]} = P \exp \left( i g \int_0^1 dt \, \mathcal{B}_i (t) \right)
\, ,
\ee
where the path is parametrized by $x_{[ii+1]} (t) =x_i - t x_{ii+1}$, $\theta_{[ii+1]} (t) = \theta_i - t \theta_{ii+1}$  and the first few terms in the superconnection 
read\footnote{These are the only components that we will need for the main calculation performed in the paper.} 
\be
\label{SuperconnectionB}
\mathcal{B}_i (t) 
= 
- \frac{1}{2} \bra{i} A (t) |i] - \frac{i}{2} \chi_i^A [\bar\psi_A (t) | i] - \frac{i}{2} \chi_i^A \left( \frac{1}{2} \bra{\theta^A_{i,i+1} (t)} D |i] + \eta_i^B \right) \bar\phi_{AB} (t)
+ \dots
\, .
\ee
The duality relation between the two objects is of the following form
\be
\vev{\mathcal{W}_{n;k} (x_i, \theta_i)} = \left( \frac{g^2 N_c}{4 \pi^2} \right)^k \widehat{\mathcal{A}}_{n;k} (\lambda_i, \widetilde\lambda_i, \eta_i)
\, .
\ee
It is a generalization of the duality for the lowest $k=0$ MHV component \cite {AldMal08,KorDruSok08,BraHesTra07,DruHenKorSok07,KorDruHenSok08} 
elucidated by now through multi-loop calculations 
\cite{ABDK03,BDS05,AnaBraHesKhoSpeTra09,DelDuhSmi9,GonSprVerVol10,BerDixKosRoiSprVerVol08,CacSprVol08}.
The above equation has the unusual property of mixing orders of perturbation theory on both sides of the equation, for instance, the $\ell$-th order $k=1$ 
NMHV amplitude emerges from a $(\ell+1)$-loop computation of the superloop, however, from terms quartic in Grassmann variables, etc.

To lowest order in coupling the amplitudes and their duals on the super Wilson loop side are expected to be invariant under the so-called dual 
superconformal symmetry which acts on the dual coordinates $(x_i, \theta_i)$ \cite{DruHenKorSok10}. At subleading orders in coupling, some of 
the symmetry generators are broken by the ultraviolet regulator in a predictable fashion. This was clearly demonstrated in great details for MHV amplitudes 
and its dual bosonic Wilson loop in Ref.\ \cite{DruHenKorSok07}. However, the first encounter with the superloop's $\eta^4-$component, dual to the tree 
NMHV amplitudes, demonstrated that the former suffer from another anomalous effect \cite{BelKorSok11}. Namely, the use of the Four-Dimensional 
Helicity scheme \cite{BerFreDixWon02}, adopted for the bulk of higher loop calculations on the amplitudes side as it preserves the spinor-helicity formalism, 
induces a conformal and supersymmetric anomaly which breaks the above correspondence \cite{BelKorSok11,Bel12}. However, this anomalous contribution 
has a universal form and can be subtracted away in a consistent manner, restoring the supersymmetry and conformal symmetry and thus resuscitating the 
conjectured duality. It is important to realize at the NMHV level, the degree four Grassmannian structure becomes anomalous provided it contains at most 
three adjacent indices \cite{Bel12}, e.g., $\chi_{i-1}^2 \chi_i \chi_{i+1}$, $\chi_{i-1}^2 \chi_{i}^2$ etc. Therefore, any structure where at least one of the indices 
is not adjacent to the rest will be conformal and given by the corresponding component of the $R$-invariants. The question still remains whether those 
components that were not anomalous at leading order develop unexpected anomalies once computed at subleading orders. This is the issue that we will 
address in the present study.

%%%%%%%%%%%%%%%%%%%%%%%%%%%%%%%%%%%%%%%%%%%%%%%%%%%%%%%%%%%%%%%%%%%%%
%            Figure
%%%%%%%%%%%%%%%%%%%%%%%%%%%%%%%%%%%%%%%%%%%%%%%%%%%%%%%%%%%%%%%%%%%%%
\begin{figure}[t]
\begin{center}
\mbox{
\begin{picture}(0,150)(100,0)
\put(0,0){\insertfig{7}{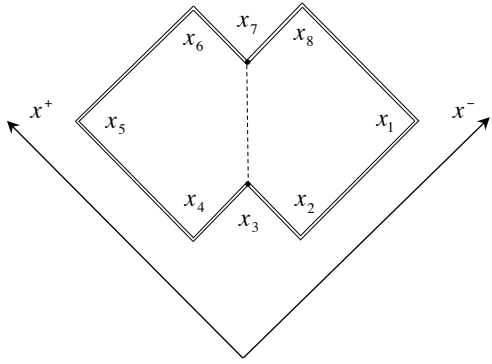}}
\end{picture}
}
\end{center}
\caption{ \label{1loop} Null octagonal Wilson loop contour in the two-dimensional kinematics. The exchanged scalar between the cusps $x_3$
and $x_7$ selects the one-loop $\chi_2 \chi_3 \chi_6 \chi_7$ component of the superloop and expressible in terms of the corresponding 
Grassmann projection of the $R_{8;37}$ superinvariant.}
\end{figure}
%%%%%%%%%%%%%%%%%%%%%%%%%%%%%%%%%%%%%%%%%%%%%%%%%%%%%%%%%%%%%%%%%%%%%

We will perform a two-loop computation of a non-anomalous component at leading order, picking $\chi_2 \chi_3 \chi_6 \chi_7$ as the object of
analysis. In order to be able to track explicitly all intricacies of cancellations between Feynman diagrams without the complications of dealing with 
higher-degree transcendental functions intrinsic to computations in the full four-dimensional kinematics, we will restrict the contour of the superloop 
to a two-dimensional subspace \cite{AldMal08,DelDuhSmi10,Bel11}. In this situation, the highest transcendentality that one can expect in the result 
is degree two, which encompasses dilogarithms and squares of logs (as well as their lower powers). Moreover, the first nontrivial loop has the octagonal 
shape as shown in Fig.\ \ref{1loop}, along with the definition of the light-like directions of the segments. The component in question of the tree NMHV 
amplitude is expressed in terms of $R_{8;37}$,
\be
R_{8;37} = \frac{\chi_2 \chi_3 \chi_6 \chi_7}{2 \vev{23} \vev{67} x_{73}^+ x_{73}^-} + \dots
\, .
\ee
This results can be easily reproduced by evaluating the corresponding component of the one-loop super Wilson loop, which is determined by the
correlation function of two superconnections $\mathcal{B}$ each of which gets reduced to boundary terms%
\footnote{For instance, $\mathcal{W}_{[32]} = \mathcal{W}_{[43]} = 1 + \ft12 g^2 \chi_2^A \chi_3^B \bar\phi_{AB} (x_3)/ \vev{23}$ and 
analogously for other two segments adjacent to the vertex $x_7$.} with the scalar field localized at the vertices $x_3$ and $x_7$, 
\be
\vev{\mathcal{W}_{8,1}^{(1)}}
= 
- \frac{g^2}{4 \pi^2} \frac{C_F}{2}  \frac{\chi_2 \chi_3 \chi_6 \chi_7}{\vev{23} \vev{67} x_{73}^+ x_{73}^-} (- \mu^2 x_{73}^+ x_{73}^-)^\varepsilon
\, .
\ee
Here we kept the regularized form of the one-loop result since it will be essential for the definition of the remainder function in the discussion
that follows. Notice that we absorbed transcendental constants into the rescaled mass parameter $2 \pi {\rm e}^{\gamma_E} \mu^2 \to \mu^2$. 
Removing the regulator, $\varepsilon \to 0$, we immediately see that the one-loop superloop is expressible in terms of the $R_{8;37}$ component of
the superconformal invariant. This is the expected result since the anomaly emerges only in adjacent Grassmann components as explained above.

The complexity level of the computation that follows is comparable to the two-loop calculation of the bosonic Wilson loop which is dual to MHV
amplitudes. Presently, the two-loop analysis yields the dual to the one-loop NMHV amplitude since we are extracting degree-four Grassmann
component. As in our previous studies \cite{BelKorSok11,Bel12} we will adopt the Four-Dimensional Helicity scheme \cite{BerFreDixWon02} to 
regularize divergences in Feynman graphs. This regularization is the closest one to the way one tackles infrared divergent scattering amplitudes. 
The details of the analysis are deferred to the Appendix.

\begin{figure}[t]
\begin{center}
\mbox{
\begin{picture}(0,90)(230,0)
\put(0,0){\insertfig{16}{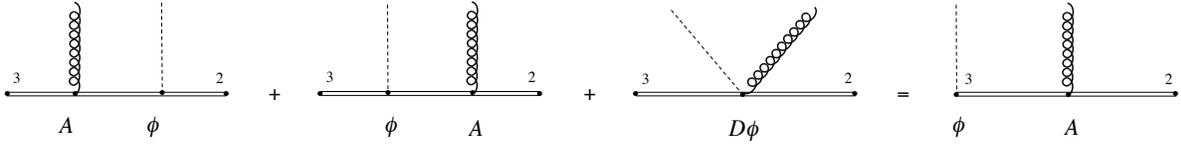}}
\end{picture}
}
\end{center}
\caption{ \label{seagullANDgluonphi} Cancellation mechanism between the different ordering of gluon and scalar emission with
the seagull terms stemming from the field component of the covariant derivative $D \bar\phi_{AB}$.}
\end{figure}

Due to the choice of the particular Grassmann component, a number of Feynman graphs should not taken into account. Namely, at second order in 
coupling, one has to include the effects from the covariant derivative (see the last term in Eq. (\ref{SuperconnectionB})) along with emission of scalar 
and gluon fields off the super-Wilson links. However, a quick inspection demonstrates that the sum of two orderings of emission along with seagull 
terms vanish as shown in Fig.\  \ref{seagullANDgluonphi}. The cancellation works as follows. Consider the $[32]$-superlink as an example. Expanding 
it to second order in $g$, and keeping track of $\chi_2 \chi_3$ component only (and ignoring fermions for a moment), we find
\ba
\mathcal{W}_{[32]} 
&\stackrel{\chi_2 \chi_3}{=}& 
\frac{i g^2}{4} \frac{\chi_2^A \chi_3^B}{\vev{23}}
\Bigg\{
\int_0^1 dt \, t \, \bra{2} [A(t), \bar\phi_{AB} (t)] |2]
\\
&-&
\int_0^1 dt \int_0^t dt' 
\left[
\bra{2} A(t) |2] \left( t' \frac{d}{d t'} + 1 \right) \bar\phi_{AB} (t')
+
\left( t \frac{d}{d t} + 1 \right) \bar\phi_{AB} (t) \bra{2} A(t') |2] 
\right]
\Bigg\}
\, . \nonumber
\ea
Here the argument of all functions involved stands for $f(t) \equiv f(x_{[23]} (t))$. The first line above displays the gauge field part of the covariant
derivative, while the terms involving derivatives in the second line emerge from its flat part. Finally, derivative-free contributions in the
integrand of the two-fold integrals come from two ordering of inserting the gluon and the scalar field into the $[32]$-link. The follow-up
simplification of this expression is straightforward and one finds that the scalar fields is nailed down to the vertex at $x = x_3$ while the gluon
is emitted from any point on the link
\be
\mathcal{W}_{[32]} 
\stackrel{\chi_2 \chi_3}{=} - \frac{i g^2}{4} \frac{\chi_2^A \chi_3^B}{\vev{23}} \bar\phi_{AB} (x_3) \int_0^1 dt \, \bra{2} A (t) |2]
\, .
\ee
It takes the form of the leading order scalar emission vertex and a bosonic Wilson segment attached to it. Analogous arguments apply with minor 
modifications to other superlinks adjacent to the cusps at $x_3$ and $x_7$ yielding contributions with the scalar localized at the cusps
and gluon strings attached to it.

%%%%%%%%%%%%%%%%%%%%%%%%%%%%%%%%%%%%%%%%%%%%%%%%%%%%%%%%%%%%%%%%%%%%%
%            Figure
%%%%%%%%%%%%%%%%%%%%%%%%%%%%%%%%%%%%%%%%%%%%%%%%%%%%%%%%%%%%%%%%%%%%%
%%%%%%%%%%%%%%%%%%%%%%%%%%%%%%%%%%%%%%%%%%%%%%%%%%%%%%%%%%%%%%%%%%%%%

\begin{figure}[t]
\begin{center}
\mbox{
\begin{picture}(0,370)(240,0)
\put(0,0){\insertfig{17}{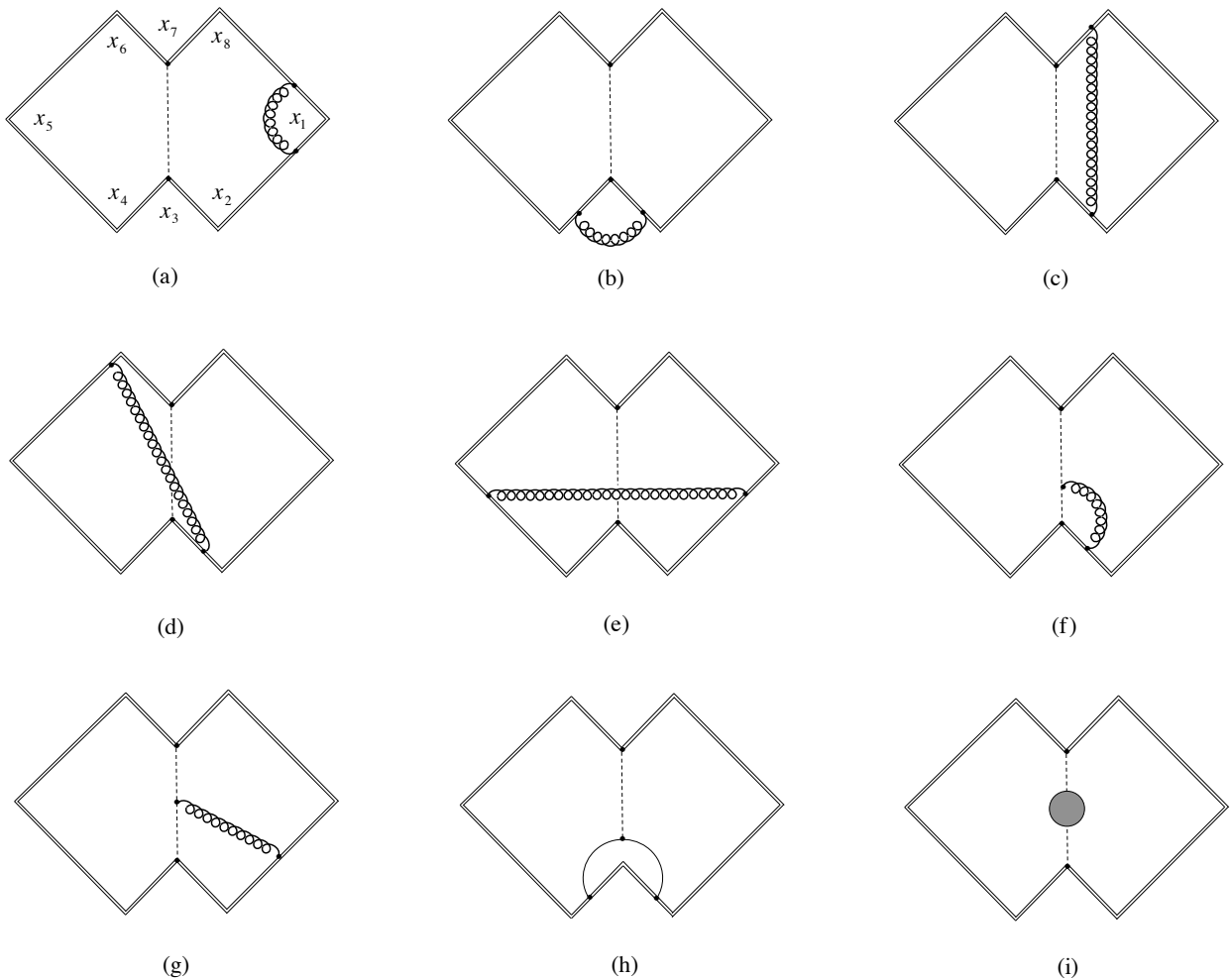}}
\end{picture}
}
\end{center}
\caption{ \label{2loopdiagrams} All topologies of Feynman diagrams contributing to the $\chi_2 \chi_3 \chi_6 \chi_7$ component of the
supersymmertic Wilson loop at two loop order. The blob on the scalar line in (i) stands for the sum of vacuum polarization bubbles due to
gauge fields and gauginos.}
\end{figure}

As a consequence of this consideration, we are left with graphs where the scalar can only spill off the cusps at $x_3$ and $x_7$, thus generating 
two-loop diagrams shown in Fig.\ \ref{2loopdiagrams} (a-g) along with other attachment of gluons to other segments of the contour to form a gauge-invariant
set.  Since the same Grassmann structures, either $\chi_2 \chi_3$ or $\chi_6 \chi_7$, can be induced by fermions emitted off the two adjacent links, to the 
the given order in coupling, there is an extra graph of the type (h). Notice that it is required by supersymmetry of the superloop and it will be instrumental for 
the cancellation of the double-pole divergences in the non-abelian color structure $C_F C_A$.

To present the result of our analysis, we will strip the dependence on the Grassmann variables and powers of the gauge coupling constant from the 
component of the super Wilson loop that we are interested in
\be
\vev{\mathcal{W}_{8;1}^{(2)}}
= 
\frac{1}{2} \left( \frac{g^2}{4 \pi^2} \right)^2 \frac{\chi_2 \chi_3 \chi_6 \chi_7}{\vev{23} \vev{67} x^+_{37} x^-_{37}}  
\sum_\alpha \left( C_F^2 w^{\scriptscriptstyle\rm A}_\alpha - \frac{1}{2} C_F C_A w^{\scriptscriptstyle\rm NA}_\alpha \right)
\, .
\ee
Here the sum runs over the diagrams displayed in Fig.\ \ref{2loopdiagrams} and split it into abelian and maximally non-abelian color Casimirs. The 
contribution to the abelian part of the Wilson loop stems from the diagrams in (a), (b), (c), (d) and (e) and the result of rather elementary computations gives
\ba
w^{\scriptscriptstyle\rm A}_{\rm(a)}
&=&
- \frac{1}{2}
(- x_{73}^+ x_{73}^- \mu^2)^\varepsilon 
\big[ 
(- x_{72}^+ x_{83}^- \mu^2)^ \varepsilon 
+ 
(- x_{72}^+ x_{87}^- \mu^2)^\varepsilon 
+
(- x_{76}^+ x_{74}^- \mu^2)^\varepsilon 
\\
&&
\qquad\qquad\qquad\quad
+
(- x_{63}^+ x_{74}^- \mu^2)^\varepsilon 
+
(- x_{36}^+ x_{34}^- \mu^2)^\varepsilon 
+
(- x_{23}^+ x_{83}^- \mu^2)^\varepsilon 
\big] 
\left( \varepsilon^{- 2} + \zeta_2 \right)
\, , \nonumber\\
\nonumber\\
w^{\scriptscriptstyle\rm A}_{\rm(c)}
&=&
- \frac{1}{2} \ln \frac{x_{83}^-}{x_{73}^-} \ln \frac{x_{73}^+}{x_{72}^+} 
- \frac{1}{2} \ln \frac{x_{74}^-}{x_{73}^-} \ln \frac{x_{73}^+}{x_{63}^+} 
\, , \\
\nonumber\\
w^{\scriptscriptstyle\rm A}_{\rm(b)}
&=&
w^{\scriptscriptstyle\rm NA}_{\rm(b)}
\, , \qquad
w^{\scriptscriptstyle\rm A}_{\rm(d)}
=
w^{\scriptscriptstyle\rm NA}_{\rm(d)}
\, , \qquad
w^{\scriptscriptstyle\rm A}_{\rm(e)}
=
w^{\scriptscriptstyle\rm NA}_{\rm(e)}
\, ,
\ea
with $w^{\scriptscriptstyle\rm NA}_{\rm(b,d,e)}$ displayed below. The abelian part of the expression has the following multiplicative structure
\be
\frac{1}{2} \left( \frac{g^2}{4 \pi^2} \right)^2 \frac{\chi_2 \chi_3 \chi_6 \chi_7}{\vev{23} \vev{67} x^+_{37} x^-_{37}}  
\sum_\alpha C_F^2 w^{\scriptscriptstyle\rm A}_\alpha 
=
\vev{\mathcal{W}_{8;1}^{(1)}}
\vev{\mathcal{W}_{8;0}^{(1)}}
\, ,
\ee
where first factor is the one-loop correction to the superloop $\vev{\mathcal{W}_{8;1}^{(1)}}$ defining the tree NMHV amplitude and the second is
the one-loop correction to the bosonic loop which is equal to $C_F \sum_\alpha w_\alpha^{\scriptscriptstyle\rm A}$ up to factors of the coupling 
constant. Thus the remainder function, defined by subtracting the ultraviolet divergent contributions  conventionally by
\be
\mathcal{R}_{8; 1}^{(2)}
=
\vev{\mathcal{W}_{8;1}^{(2)}}
-
\vev{\mathcal{W}_{8;1}^{(1)}} \vev{\mathcal{W}_{8;0}^{(1)}}
\, ,
\ee
is solely defined by the maximally nonabelian color. Therefore, the sum of all corresponding contributions has to be dual conformally invariant.

The analysis of the maximally non-abelian contributions is more involved. The result of rather lengthy calculations can be cast to the following form
\begin{align}
w^{\scriptscriptstyle\rm NA}_{\rm(b)}
&=
- \frac{1}{2} (- x_{73}^+ x_{73}^- \mu^2)^\varepsilon \left[ (- x_{23}^+ x_{34}^- \mu^2)^ \varepsilon + (- x_{76}^+ x_{87}^- \mu^2)^\varepsilon \right] \left( \varepsilon^{- 2} + \zeta_2 \right)
\, , \\[3mm]
w^{\scriptscriptstyle\rm NA}_{\rm(d)}
&= 
\frac{1}{2}
\ln \frac{x_{72}^+}{x_{62}^+} \ln \frac{x_{73}^-}{x_{78}^-}
+
\frac{1}{2}
\ln \frac{x_{76}^+}{x_{73}^+} \ln \frac{x_{84}^-}{x_{74}^-}
+
\frac{1}{2}
\ln \frac{x_{63}^+}{x_{62}^+} \ln \frac{x_{73}^-}{x_{34}^-}
+
\frac{1}{2}
\ln \frac{x_{23}^+}{x_{73}^+} \ln \frac{x_{84}^-}{x_{83}^-}
\, , \\
\nonumber\\
w^{\scriptscriptstyle\rm NA}_{\rm(e)}
&= 
\frac{1}{2}
\ln \frac{x_{62}^+}{x_{67}^+} \ln \frac{x_{78}^-}{x_{48}^-}
+
\frac{1}{2}
\ln \frac{x_{62}^+}{x_{23}^+} \ln \frac{x_{34}^-}{x_{84}^-}
\, , \\[3mm]
w^{\scriptscriptstyle\rm NA}_{\rm(f)}
&= 
\varepsilon^{-1} (- x_{73}^+ x_{73}^- \mu^2)^{2 \varepsilon}
\bigg[
\frac{1}{2}
\left[ 1 - \frac{x_{73}^+}{2 x_{23}^+} \right] \ln \frac{x_{72}^+}{x_{73}^+} 
+ 
\frac{1}{2}
\left[ 1 - \frac{x_{73}^+}{2x_{76}^+} \right] \ln \frac{x_{63}^+}{x_{73}^+}
\\
&
\qquad\qquad\qquad\quad \,
+  
\frac{1}{2}
\left[ 1 + \frac{x_{73}^-}{2 x_{87}^-} \right] \ln \frac{x_{83}^-}{x_{73}^-} 
+
\frac{1}{2}
\left[ 1+ \frac{x_{73}^-}{x_{34}^-} \right] \ln \frac{x_{74}^-}{x_{73}^-}
+
2
\bigg]
\nonumber\\
&+
\frac{x_{73}^+}{2x_{23}^+} {\rm Li}_2 \left( \frac{x_{23}^+}{x_{73}^+} \right)
+
\frac{x_{73}^+}{2x_{76}^+} {\rm Li}_2 \left( \frac{x_{76}^+}{x_{73}^+} \right)
-
\frac{x_{73}^-}{2x_{34}^-} {\rm Li}_2 \left( \frac{x_{43}^-}{x_{73}^-} \right)
-
\frac{x_{73}^-}{2x_{87}^-} {\rm Li}_2 \left( \frac{x_{78}^-}{x_{73}^-} \right)
- 2
\nonumber\\
&+
\frac{1}{4} \left[ 1- \frac{x_{73}^+}{2 x_{23}^+} \right] \ln^2 \frac{x_{72}^+}{x_{73}^+} 
+
\frac{1}{4} \left[ 1 - \frac{x_{73}^+}{2x_{76}^+} \right] \ln^2 \frac{x_{63}^+}{x_{73}^+}
\nonumber\\
&
\qquad\qquad\qquad\qquad\quad \ 
+ 
\frac{1}{4} \left[ 1 + \frac{x_{73}^-}{2x_{87}^-} \right] \ln^2 \frac{x_{83}^-}{x_{73}^-} 
+
\frac{1}{4} \left[ 1 + \frac{x_{73}^-}{2x_{34}^-} \right] \ln^2 \frac{x_{74}^-}{x_{73}^-}
\, , \nonumber\\[3mm]
\label{w-g}
w^{\scriptscriptstyle\rm NA}_{\rm(g)}
&=
\varepsilon^{- 1} (- x_{73}^+ x_{73}^- \mu^2)^{\varepsilon}
\bigg[
\frac{1}{4} (- x_{72}^+ x_{73}^- \mu^2)^{\varepsilon} \frac{x_{73}^+}{x_{23}^+} \ln \frac{x_{72}^+}{x_{73}^+}
+
\frac{1}{4} (- x_{63}^+ x_{73}^- \mu^2)^{\varepsilon} \frac{x_{73}^+}{x_{76}^+} \ln \frac{x_{63}^+}{x_{73}^+}
\\
&\qquad\qquad\qquad\quad
-
\frac{1}{4} (- x_{73}^+ x_{83}^- \mu^2)^{\varepsilon} \frac{x_{73}^-}{x_{87}^-} \ln \frac{x_{83}^-}{x_{73}^-}
-
\frac{1}{4} (- x_{73}^+ x_{74}^- \mu^2)^{\varepsilon} \frac{x_{73}^-}{x_{34}^-} \ln \frac{x_{74}^-}{x_{73}^-}
\bigg]
\nonumber\\
&-
\frac{1}{8} \frac{x_{73}^+}{x_{23}^+} \ln^2 \frac{x_{72}^+}{x_{73}^+}
-
\frac{1}{8} \frac{x_{73}^-}{x_{78}^-} \ln^2 \frac{x_{83}^-}{x_{73}^-}
+
\frac{1}{8} \frac{x_{73}^+}{x_{67}^+} \ln^2 \frac{x_{63}^+}{x_{73}^+}
+
\frac{1}{8} \frac{x_{73}^-}{x_{34}^-} \ln^2 \frac{x_{74}^-}{x_{73}^-}
-
\zeta_2
\nonumber\\
&-
\frac{1}{2}
\ln \frac{x_{23}^+ x_{83}^-}{x_{73}^+ x_{73}^-} \ln \frac{x_{73}^+ x_{78}^-}{x_{72}^+ x_{73}^-}
+
\frac{1}{2}
\ln \frac{x_{36}^+ x_{34}^-}{x_{73}^+ x_{73}^-} \ln \frac{x_{73}^+ x_{47}^-}{x_{67}^+ x_{73}^-}
+
\ln\frac{x_{38}^-}{x_{73}^-} \ln\frac{x_{78}^-}{x_{73}^-}
+
\ln\frac{x_{63}^+}{x_{73}^+} \ln\frac{x_{67}^+}{x_{73}^+}
\nonumber\\
&+
\left[ 1 - \frac{x_{73}^+}{2x_{23}^+}  \right] {\rm Li}_2 \left( \frac{x_{23}^+}{x_{73}^+} \right)
+
\left[ 1 - \frac{x_{73}^+}{2x_{76}^+} \right] {\rm Li}_2 \left( \frac{x_{76}^+}{x_{73}^+} \right)
\nonumber\\
&
\qquad\qquad\qquad\qquad\qquad 
+
\left[ 1 + \frac{x_{73}^-}{2x_{87}^-}  \right] {\rm Li}_2 \left( \frac{x_{78}^-}{x_{73}^-} \right)
+
\left[ 1+ \frac{x_{73}^-}{2x_{34}^-} \right] {\rm Li}_2 \left( \frac{x_{43}^-}{x_{73}^-} \right)
\, , \nonumber\\[3mm]
\label{w-h}
w^{\scriptscriptstyle\rm NA}_{\rm(h)}
&= 
\frac{1}{2} (- x_{73}^+ x_{73}^- \mu^2)^\varepsilon (- x_{23}^+ x_{34}^- \mu^2)^ \varepsilon  
\left[ \varepsilon^{- 2} + \zeta_2 - \varepsilon^{-1} \ln\frac{x_{72}^+ x_{74}^-}{x_{73}^+ x_{73}^-} \right]
\\
&+
\frac{1}{2} (- x_{73}^+ x_{73}^- \mu^2)^\varepsilon (- x_{76}^+ x_{87}^- \mu^2)^\varepsilon
\left[ \varepsilon^{- 2} + \zeta_2 -  \varepsilon^{-1} \ln\frac{x_{63}^+ x_{83}^-}{x_{73}^+ x_{73}^-} \right]
\nonumber\\
&-
\frac{1}{4} \ln^2 \frac{x_{72}^+ x_{73}^-}{x_{73}^+ x_{74}^-} - \frac{1}{4} \ln^2 \frac{x_{63}^+ x_{73}^-}{x_{73}^+ x_{83}^-}
-
{\rm Li}_2 \left( \frac{x_{23}^+}{x_{73}^+} \right)
-
{\rm Li}_2 \left( \frac{x_{43}^-}{x_{73}^-} \right)
-
{\rm Li}_2 \left( \frac{x_{87}^+}{x_{73}^+} \right)
-
{\rm Li}_2 \left( \frac{x_{76}^+}{x_{73}^+} \right)
\, , \nonumber\\[3mm]
w^{\scriptscriptstyle\rm NA}_{\rm(i)}
&= 
- \varepsilon^{-1} (- x_{73}^+ x_{73}^- \mu^2)^{2 \varepsilon}  - 2
\, .
\end{align}

%%%%%%%%%%%%%%%%%%%%%%%%%%%%%%%%%%%%%%%%%%%%%%%%%%%%%%%%%%%%%%%%%%%%%
%            Figure
%%%%%%%%%%%%%%%%%%%%%%%%%%%%%%%%%%%%%%%%%%%%%%%%%%%%%%%%%%%%%%%%%%%%%
%%%%%%%%%%%%%%%%%%%%%%%%%%%%%%%%%%%%%%%%%%%%%%%%%%%%%%%%%%%%%%%%%%%%%

Summing up the diagrams, we find that, as anticipated, that all ultraviolet poles chancel among different graphs in the remainder function. In particular, 
the fermionic graph (h) exactly cancels the maximally non-abelian color structure in diagram (b). The sum of single poles vanish as well provided we set 
$(- x_{ij}^+ x_{kl}^- \mu^2)^\varepsilon \to 1$, but otherwise they induce finite contributions. The latter are of paramount importance in removing all 
terms proportional to squares of the logs that depend solely either on plus or minus components of the dual coordinates. Finally, all dilogarithms present 
in individual graphs sum-up to zero in the remainder function as well. As a result the latter takes a very simple form
\ba
\mathcal{R}_{8; 1}^{(2)}
\!\!\!&=&\!\!\!
\left( \frac{g^2}{4 \pi^2} \right)^2 \frac{C_F C_A}{8} \frac{\chi_2 \chi_3 \chi_6 \chi_7}{\vev{23} \vev{67} x_{73}^+ x_{73}^-}
\\
&\times&\!\!\!
\left[ \, 
\ln u^+ \ln u^- + \ln u^+ \ln (1+ u^-) + \ln (1 + u^+) \ln u^- - \ln (1 + u^+) \ln (1+u^-)
+ 2 \zeta_2 \, 
\right]
\, , \nonumber
\ea
upon the introduction of the conformal cross-ratios
\be
u^+ = \frac{x^+_{32} x^+_{67}}{x^+_{62} x^+_{73}}
\, , \qquad
u^- = \frac{x^-_{87} x^-_{34}}{x^-_{84} x^-_{73}}
\, .
\ee
This expression is an agreement with the result of a recent analysis that bypasses the calculation of the Feynman graphs and finds the result in
question by integrating the Ward identities associated with $\bar{Q}$ supersymmetry from the the tree-level NNMHV amplitude \cite{CarHe11}.

Presently, we verified by a brute-force Feynman graph calculation the duality between the supersymmetric extension of the null polygonal Wilson 
loop and the superamplitude in maximally supersymmetric gauge theory. Our analysis elucidates the validity of the correspondence for Grassmann
components which do not involve at least three adjacent particle indices. The latter were shown to be anomalous already at one-loop order.
However, once the universal conformal anomaly is subtracted out, the duality gets restored. In the forthcoming work \cite{BelCar12}, we will demonstrate 
how one can perform a super-gauge transformation on the super Wilson loop in order to gauge away in a systematic manner the notorious anomalous 
contributions.

\vspace{5mm}

We would like to thank Gregory Korchemsky for usefull correspondence and instructive comments and Song He for making the component form of NMHV 
superamplitudes in restricted kinematics available to us. This work was supported by the U.S. National Science Foundation under grants PHY-0757394 
and PHY-1068286.

\appendix
\section{Details of the calculation}

The bulk of graphs is easy to compute. Let us pay special attention to a couple of them that are not as straightforward, namely diagrams (g) and (h)
in Fig.\ \ref{2loopdiagrams}.

\subsection{Diagram (g)}

Using the usual Feynman rules, we find the following integral representation for the graph (g),
\ba
\mathcal{W}^{(2)}_{8;1 (h)} 
&=& 
i \frac{g^4  C_F C_A}{(4 \pi^{2 - \varepsilon})^3} \, x_{12}^- \, \frac{\chi_2 \chi_3 \chi_6 \chi_7}{\vev{23} \vev{67}}
\Gamma^2 (1 - \varepsilon) \Gamma (2 - \varepsilon)
\int_0^1 dt \, J_g (t) \, , \nonumber
\ea
where
\be
J_g (t) = \int d^{4 - 2 \varepsilon} z \, \frac{1}{[-(z - x_{[12]}(t))^2]^{1 - \varepsilon} [-(z - x_3)^2]^{1 - \varepsilon} [-(z - x_7)^2]^{1 - \varepsilon}}
\left[
\frac{(z - x_7)^+}{(z - x_7)^2}
-
\frac{(z - x_3)^+}{(z - x_3)^2}
\right]
\, .
\ee
As a first step, we use the Feynman parametrization to put $J_g$ into the form
\ba
J_g (t)
&=&
- \frac{i \pi^{2 - \varepsilon} \Gamma (2 - 2\varepsilon)}{\Gamma^2 (1-\varepsilon) \Gamma(2-\varepsilon)} 
\\
&\times&
\mu^{4 \varepsilon}
\int_0^1
\frac{ds_1 \, ds_2 \, ds_3 \, \delta \left( \sum_{i = 1}^3 s_i - 1 \right) (s_1 s_2 s_3)^{- \varepsilon}
}{
\big[- s_1 s_2 x_{[12]7}^2 - s_1 s_3 x_{[12]3}^2 - s_2 s_3 x_{73}^2 \big]^{2 - 2 \varepsilon}}
\left[
s_1 s_3 x_{[12]3}^+ - s_1 s_2 x_{[12]7}^+ + 2 s_2 s_3 x_{73}^+
\right]
\, . \nonumber
\ea
Now,  expressing $s_3$ in terms of the other two variables via the $\delta-$function constraint and changing the integration variables as\footnote{Here and 
below $\bar{s} \equiv 1 - s$ for any variables.}  $s_2 \to \bar{s}_1 s_2$ and as a consequence $s_3 \to \bar{s}_1 \bar{s}_2$, we can integrate with respect to $t$ 
to get
\ba
&&
\int_0^1 dt J_g (t)
=
\frac{i \pi^{2 - \varepsilon} \Gamma (1 - 2\varepsilon)}{\Gamma^2 (1 - \varepsilon) \Gamma (2 - \varepsilon) }
\frac{(- 2\mu^2)^{2 \varepsilon}}{4 x_{12}^-}
\int_0^1 \frac{ds_1}{s_1^{1 + \varepsilon}} \int_0^1 \frac{ds_2}{(s_2 \bar{s}_2)^\varepsilon}
\frac{s_1 \bar{s}_2 x_{23}^+ - s_1 s_2 x_{17}^+ + 2 \bar{s}_1 s_2 \bar{s}_2 x_{73}^+}{s_2 x_{17}^+ + \bar{s}_2 x_{23}^+}
\nonumber\\
&&
\qquad\quad\times
\bigg[
\frac{1}{[ s_1 s_2 x_{71}^+ x_{71}^- + s_1 \bar{s}_2 x_{12}^- x_{23}^+ + \bar{s}_1 s_2 \bar{s}_2 x_{73}^+ x_{73}^- ]^{1 - 2 \varepsilon}}
-
\frac{1}{s_2^{1 - 2\varepsilon} (x_{73}^-)^{1 - 2 \varepsilon}}
\frac{1}{[s_1 x_{71}^+ + \bar{s}_1 \bar{s}_2 x_{73}^+]^{1 - 2 \varepsilon}}
\bigg]
. \nonumber\\
\ea
It might appear that there is a double pole emerging from this integrals, however, the only pole that realizes is from the $s_2$
integration in the vicinity of $s_2 = 0$, while the expression in curly brackets scales as $s_1$ and it tends to zero canceling the
potential singular behavior. We extract the pole in the $s_2$ integral via the following formula
\be
\int_0^1 \frac{ds_2}{s_2^{1 - \varepsilon}} f(s_2) = \varepsilon^{-1} f(0) + \int_0^1 \frac{ds_2}{s_2^{1 - \varepsilon}} [f(s_2) - f(0)]
\, .
\ee
This allows us to cast the result after some manipulations into the form
\ba
\int_0^1 dt J_g (t)
= 
-
\frac{i \pi^{2 - \varepsilon \Gamma (1 - 2\varepsilon)}}{4 \Gamma^2(1-\varepsilon) \Gamma(2-\varepsilon)} \frac{1}{x_{12}^- x_{73}^-}
\left[
\varepsilon^{-1} (- 2\mu^2 x_{73}^-)^{2\varepsilon} I_1 + \left( x_{73}^- - 2 x_{71}^- \right) I_2 + x_{73}^+ I_3
\right]
\, ,
\ea
with the set of $I_i$ integrals that can be evaluated with the result
\ba
I_1 
&=&
\int_0^1 \frac{ds_1}{s_1^\varepsilon} \frac{1}{[ x_{71}^+ s_1 + x_{73}^+ \bar{s}_1]^{1 - 2\varepsilon}}
\nonumber\\
&=& (- 2 \mu^2 x_{73}^-)^{-2 \varepsilon}
\frac{1}{x_{23}^+}
\left[
(- 2 \mu^2 x_{71}^+ x_{73}^-)^{\varepsilon} (- 2 \mu^2 x_{73}^+ x_{73}^-)^{\varepsilon} \ln \frac{x_{71}^+}{x_{73}^+}
-
\varepsilon {\rm Li}_2 \left( \frac{x_{23}^+}{x_{73}^+} \right)
\right]
\, , \\
I_2 &=& 
\int_0^1 ds_1 ds_2 \, \frac{1}{ x_{71}^+ x_{71}^- s_1 s_2 + x_{23}^+ x_{12}^- s_1 \bar{s}_2 + x_{73}^+ x_{73}^- \bar{s}_1 s_2 \bar{s}_2}
\nonumber\\
&=&
\frac{1}{x_{71}^+ x_{73}^- - x_{73}^+ x_{71}^-}
\left[
\ln\frac{x_{73}^+ x_{71}^-}{x_{71}^+ x_{73}^-} \ln\frac{x_{23}^+ x_{12}^-}{x_{73}^+ x_{73}^-}
+ 2 {\rm Li}_2 \left( \frac{x_{12}^-}{x_{73}^-} \right)
- 2 {\rm Li}_2 \left( \frac{x_{23}^-}{x_{73}^+} \right)
\right]
\, , \\
I_3 &=& 
\int_0^1 ds_1 ds_2 \, \frac{\bar{s}_1}{ [ x_{71}^+ s_1 + x_{73}^+ \bar{s}_1] [x_{71}^+ s_1 + x_{73}^+ \bar{s}_1\bar{s}_2]}
=
- \frac{1}{x_{23}^+ x_{73}^+}
\left[
\ln^2\frac{x_{71}^+}{x_{73}^+} +  2 {\rm Li}_2 \left( \frac{x_{23}^+}{x_{73}^+} \right)
\right]
\, .
\ea
Notice that the second integral involves a denominator that mixes both plus and minus components. This effects disappears one we add a mirror
symmetric diagram yielding a factorized product of function of plus and minus variables. Summing all diagrams of this topology we get the expression 
in Eq.\ (\ref{w-g}).

\subsection{Diagram (h)}

Now, we turn to the second graph. 
\ba
\mathcal{W}^{(2)}_{8;1 (h)} 
&=& 
- i \frac{g^4  C_F C_A}{(4 \pi^{2 - \varepsilon})^3} \chi_2 \chi_3 \chi_6 \chi_7 \frac{[23]}{\vev{67}}
\Gamma (1 - \varepsilon) \Gamma^2 (2 - \varepsilon)
\int_0^1 ds \int_0^1 dt \, J_h (t, s) \, , \nonumber
\ea
where we have used the identity $[2| (z - x_{23} (t)) (z - x_{[34]} (s)) |3] = [23] (z - x_3)^2$ in order to define the coordinate integral
\be
J_h (t, s) 
=
\mu^{4 \varepsilon} \int d^{4 - 2 \varepsilon} z
\frac{(z - x_3)^2}{[-(z - x_7)^2]^{1 - \varepsilon} [-(z - x_{[23]} (t))^2]^{2 - \varepsilon} [-(z - x_{[34]} (s))^2]^{2 - \varepsilon}}
\, .
\ee
By means of the standard Feynman parametrization, one can cast it in the form after integration over $z$
\be
J_h (t, s) = - \frac{i \pi^{2 - \varepsilon} \Gamma (2 - 2\varepsilon)}{\Gamma (1 - \varepsilon) \Gamma^2 (2 - \varepsilon)}
\big[
4 (1 - \varepsilon) \, x_{73}^+ x_{73}^- \, I_1 (t, s) - I_2 (t, s)
\big]
\, ,
\ee
where
\ba
I_1 (t, s) &=& \mu^{4 \varepsilon} \int_0^1 \frac{ds_1 \, ds_2 \, ds_3 \, \delta \left( \sum_{i = 1}^3 s_i - 1 \right) (s_1 s_2 s_3)^{1 - \varepsilon}
}{
\big[- s_1 s_2 x_{7[23]}^2 - s_1 s_3 x_{7[34]}^2 - s_2 s_3 x_{[23][34]}^2 \big]^{3 - 2 \varepsilon}}
\, , \\
I_2 (t, s) &=& \varepsilon \, \mu^{4 \varepsilon} \int_0^1 \frac{ds_1 \, ds_2 \, ds_3 \, \delta \left( \sum_{i = 1}^3 s_i - 1 \right) s_1^{- \varepsilon} (s_2 s_3)^{1 - \varepsilon}
}{
\big[- s_1 s_2 x_{7[23]}^2 - s_1 s_3 x_{7[34]}^2 - s_2 s_3 x_{[23][34]}^2 \big]^{2 - 2 \varepsilon}}
\, .
\ea 
To perform the integrations efficiently, we remove the $s_3$ variable with the $\delta$-function and then rescale $s_2 \to \bar{s}_1 s_2$ which implies 
$s_3 \to \bar{s}_1 \bar{s}_2$. The denominator admits a factorized form with two factors both linear in $s_1$, $s_1 s_2 x_{7[23]}^2 + s_1 s_3 x_{7[34]}^2 
+ s_2 s_3 x_{[23][34]}^2 = 2 \bar{s}_1 \left[  A s_1 + B \right]$ with
\be
A = (x_{73}^+ - \bar{t} s_2 x_{23}^+)  (x_{73}^- + s \bar{s}_2 x_{34}^-)
\, , \qquad
B = s_2 \bar{s}_2 \bar{t} s \, x_{23}^+ x_{34}^-
\, .
\ee
The next integration to be performed is with respect to $s_1$, which produces a hypergeometric functions ${_2F_1}$. However, we notice that the small-$s_1$
region yields a contribution inverse in $B$ such that the two subsequent $s$- and $t$-integration induce divergencies. This allows us to extract the leading 
inverse-power behavior of the regularized integral at small $B$ and then resum the rest for $\varepsilon = 0$. This gives
\be
\label{Ints1}
\int_0^1 \frac{ds_1 \, s_1^{1 - \varepsilon}}{[A s_1 + B]^{3 - 2 \varepsilon}} 
= 
\frac{\Gamma (1 - \varepsilon) \Gamma (2 - \varepsilon)}{\Gamma (3 - 2\varepsilon)} \frac{1}{B^{1 - \varepsilon} A^{2 - \varepsilon}}
-
\frac{1}{2 A (A + B)^2} - \frac{1}{2 A^2 (A + B)}
+
\mathcal{O} (\varepsilon)
\, .
\ee
Then integrating over the $s$ and $t$ variables, we get the following representation for the integral
\ba
\int_0^1 dt \int_0^1 ds \, I_1 (t, s) 
&=& 
\frac{1}{16} \int_0^1 dt \int_0^1 ds \int_0^1 ds_2 \, s_2 \bar{s}_2  \left[ \frac{1}{A (A + B)^2} - \frac{1}{A^2 (A + B)} \right]
\\
&-&
\frac{(- 2 \mu^2 x_{23}^+ x_{34}^-)^{\varepsilon} (- 2 \mu^2 x_{73}^+ x_{73}^-)^{\varepsilon}
}{8 \, x_{23}^+ x_{34}^- (x_{73}^+ x_{73}^-)^2} 
\frac{\Gamma (1 - \varepsilon) \Gamma (2 - \varepsilon)}{\Gamma (3 - 2\varepsilon)} 
\int_0^1 ds_2 S_\varepsilon (s_2) T_\varepsilon (s_2)
\, , \nonumber
\ea
where the divergent one-dimensional contributions are given by
\be
S_\varepsilon (s_2) 
=
\int_0^1 \frac{ds}{s^{1 - \varepsilon}} \frac{1}{[ 1 + s \bar{s}_2 \, x_{34}^-/x_{73}^- ]^{2 - \varepsilon}}
\, , \qquad
T_\varepsilon (s_2)
=
\int_0^1 \frac{dt}{\bar{t}^{1 - \varepsilon}} \frac{1}{[ 1 - \bar{t} s_2 \, x_{23}^+ /x_{73}^+]^{2 - \varepsilon}}
\, .
\ee
Their $\varepsilon-$expansion is easy to construct and reads to order $\mathcal{O} (\varepsilon)$, e.g., for $T_\varepsilon (s_2)$
\ba
\label{Tepsilon}
T_\varepsilon (s_2) 
\!\!\!&=&\!\!\! 
\frac{1}{\varepsilon} + \frac{x_{23}^+ s_2}{x_{73}^+ - x_{23}^+ s_2} - \ln\left( 1 - s_2 \frac{x_{23}^+}{x_{73}^+} \right)
\\
&+&\!\!\!
\varepsilon
\left[
\frac{x_{23}^+ s_2}{x_{73}^+ - x_{23}^+ s_2}
+
\frac{2 x_{73}^+ - x_{23}^+ s_2}{x_{73}^+ - x_{23}^+ s_2}  \ln\left( 1 - s_2 \frac{x_{23}^+}{x_{73}^+} \right)
-
\frac{1}{2} \ln^2 \left( 1 - s_2 \frac{x_{23}^+}{x_{73}^+} \right) - 2 {\rm Li}_2 \left( s_2  \frac{x_{23}^+}{x_{73}^+} \right)
\right]
\, , \nonumber
\ea
and the one for $S_\varepsilon (s_2)$ being analogous with the obvious substitutions of the defining variables. With poles
being extracted explicitly, the remaining integrations can be performed with Mathematica. The output is given
however, in a form that involves dilogarithms with arguments depending of products of plus and minus variables. Instead
of relying on known identities between the dilogarithms to simplify the result and cast it as sum of functions depending
either on plus or minus variables, the use of the formalism of symbols \cite{Gon08,GonSprVerVol10} becomes very instrumental 
for fast and efficient derivations of the sought identities. Just to give an example, we encounter the following combination 
of dilogarithms in the output,
\be
L (u, v) = {\rm Li}_2 \left( 1 + \frac{v}{u \bar{v}} \right) - {\rm Li}_2 \left( 1 + \frac{v \bar{u}}{u} \right) - {\rm Li}_2 \left( \bar{u} \bar{v} \right)
\, ,
\ee
where $u = x_{23}^+/x_{73}^+$ and $v = x_{43}^-/x_{73}^-$. In order to disentangle the $u$ and $v$ dependence, we calculate 
the symbol of the right-hand side of this identity and find after simple manupulations
\ba
\mathcal{S} \left[ L (u, v) \right] 
&=& 
- \frac{v}{u \bar{v}} \otimes \left( 1 + \frac{v}{u \bar{v}} \right) + \frac{v \bar{u}}{u} \otimes \left( 1 + \frac{v \bar{u}}{u} \right)
+ (u+ v \bar{u}) \otimes (\bar{u}\bar{v})
\\
&=& 
(u+ v \bar{u}) \otimes (\bar{u}\bar{v}) + (\bar{u}\bar{v})  \otimes (u+ v \bar{u}) - \bar{v} \otimes u - u \otimes \bar{v} - \bar{v} \otimes \bar{v}
- \bar{u} \otimes u + v \otimes \bar{v}
\, . \nonumber
\ea
From here, we can immediately read off the expression for the function itself (with a potentially present additive transcendental constant fixed by 
comparing both sides of the equation numerically),
\be
L (u, v) = \ln (\bar{u} \bar{v}) \ln (u+ v \bar{u}) - \ln\bar{v} \ln u - \frac{1}{2} \ln^2 \bar{v} + {\rm Li}_2 (u) - {\rm Li}_2 (\bar{v})
\, .
\ee
The same technique is applicable to all other terms. The sum of all terms yields expressions with arguments being functions of either $u$ or $v$ 
variables separately.

Finally, we get for the integral $I_1$,
\ba
\int_0^1 dt \int_0^1 ds \, I_1 (t, s) 
\!\!\!&=&\!\!\!
- \frac{(- 2 \mu^2 x_{23}^+ x_{34}^-)^{\varepsilon} (- 2 \mu^2 x_{73}^+ x_{73}^-)^{\varepsilon}
}{8 \, x_{23}^+ x_{34}^- (x_{73}^+ x_{73}^-)^2} 
\frac{\Gamma (1 - \varepsilon) \Gamma (2 - \varepsilon)}{\Gamma (3 - 2\varepsilon)} 
\left[
\frac{1}{\varepsilon^2} - \frac{1}{\varepsilon} \ln \frac{x_{72}^+}{x_{73}^+} - \frac{1}{\varepsilon} \ln \frac{x_{74}^-}{x_{73}^-}
\right]
\nonumber\\
&+&
\frac{1}{8 \, x_{23}^+ x_{34}^- (x_{73}^+ x_{73}^-)^2} 
+
\frac{\ln (x_{72}^+/x_{73}^+)}{16  \, (x_{23}^+)^2 x_{34}^- x_{73}^+ (x_{73}^-)^2}
-
\frac{\ln (x_{74}^-/x_{73}^-)}{16  \, x_{23}^+ (x_{34}^-)^2 (x_{73}^+)^2 x_{73}^-}
\nonumber\\
&+&
\frac{\ln^2 (x_{72}^+ x_{73}^-/x_{73}^+ x_{34}^-)}{32  \, x_{23}^+ x_{34}^- (x_{73}^+ x_{73}^-)^2}
+
\frac{{\rm Li}_2 (x_{23}^-/x_{73}^+)}{8  \, x_{23}^+ x_{34}^- (x_{73}^+ x_{73}^-)^2}
+
\frac{{\rm Li}_2 (x_{43}^-/x_{73}^-)}{8  \, x_{23}^+ x_{34}^- (x_{73}^+ x_{73}^-)^2}
\, .
\ea

The calculation of the second contribution $I_2$ is much simpler since all one is after is the double and single pole part of the integral since 
they get compensated by the overall factor of $\varepsilon$. Then in the analogous to Eq.\ (\ref{Ints1}) integral with respect to $s_1$, one has 
to keep only the first term. The subsequent integrations over $s$ and $t$ like done above in Eq.\ (\ref{Tepsilon}), immediately gives the final
answer
\be
\int_0^1 dt \int_0^1 ds \, I_2 (t, s) 
=
\frac{(- 2 \mu^2 x_{23}^+ x_{34}^-)^{\varepsilon} (- 2 \mu^2 x_{73}^+ x_{73}^-)^{\varepsilon}}{4 \, x_{23}^+ x_{34}^- x_{73}^+ x_{73}^-} 
\frac{\Gamma^2 (1 - \varepsilon)}{\Gamma (2 - 2\varepsilon)} 
\left[
\frac{1}{\varepsilon} - \frac{x_{72}^+}{x_{23}^+} \ln \frac{x_{72}^+}{x_{73}^+} + \frac{x_{74}^-}{x_{43}^-} \ln \frac{x_{74}^-}{x_{73}^-}
\right]
\, .
\ee
Summing both contributions together, we find half of the result displayed in Eq.\ (\ref{w-h}). The other half is given by the mirror symmetric diagram,
computed via the formalism outlined above.

%%%%%%%%%%%%%%%%%%%%%%%%%%%%%%%%%%%%%%%%%%%%%%%%%%%%%%%%%%%%%%%%%%%%%

\end{document}